\documentclass[sigconf]{acmart}
\AtBeginDocument{%
  \providecommand\BibTeX{{%
    \normalfont B\kern-0.5em{\scshape i\kern-0.25em b}\kern-0.8em\TeX}}}

\usepackage{multirow}
\usepackage{graphicx}
\usepackage{subcaption}
\usepackage[most]{tcolorbox} 
\usepackage{xcolor} 
\usepackage{listings} 
\usepackage{array}

\definecolor{promptbackground}{HTML}{F7F7F9} 
\definecolor{promptframe}{HTML}{D0D0D0} 
\definecolor{textgray}{HTML}{333333} 

\newtcolorbox{promptbox}{
    colback=promptbackground,
    colframe=promptframe,
    boxrule=0.25pt,
    arc=4pt,
    left=6pt,
    right=6pt,
    top=6pt,
    bottom=6pt,
    boxsep=2pt,
    breakable,
    fontupper=\ttfamily\color{textgray}\small
}

\setcopyright{none}
\copyrightyear{}

\renewcommand\footnotetextcopyrightpermission[1]{} 

\acmConference[]{}
\acmYear{}
\acmDOI{}

\begin{document}

\title{PersonaRAG: Enhancing Retrieval-Augmented Generation Systems with User-Centric Agents}

\author{Saber Zerhoudi}
\email{saber.zerhoudi@uni-passau.de}
\orcid{0000-0003-2259-0462}
\affiliation{%
  \institution{University of Passau}
  \city{Passau}
  \country{Germany}
}

\author{Michael Granitzer}
\email{michael.granitzer@uni-passau.de}
\orcid{0000-0003-3566-5507}
\affiliation{%
  \institution{University of Passau}
  \city{Passau}
  \country{Germany}
}

\renewcommand{\shortauthors}{Zerhoudi et al.}

\begin{abstract}
Large Language Models (LLMs) struggle with generating reliable outputs due to outdated knowledge and hallucinations. Retrieval-Augmented Generation (RAG) models address this by enhancing LLMs with external knowledge, but often fail to personalize the retrieval process. This paper introduces PersonaRAG, a novel framework incorporating user-centric agents to adapt retrieval and generation based on real-time user data and interactions. Evaluated across various question answering datasets, PersonaRAG demonstrates superiority over baseline models, providing tailored answers to user needs. The results suggest promising directions for user-adapted information retrieval systems. Findings and resources are available at~\textcolor{blue}{~\url{https://github.com/padas-lab-de/PersonaRAG}}.
\end{abstract}

\ccsdesc[500]{Information systems~Personalization}
\ccsdesc[500]{Information systems~Presentation of retrieval results}
\ccsdesc[300]{Information systems~Language models}

\keywords{User interactions, Retrieval-Augmented Generation (RAG), Personalized Information Retrieval, Multi-Agent RAG.}

\maketitle

\begin{figure}[h]
   \centering
   \includegraphics[width=0.8\linewidth]{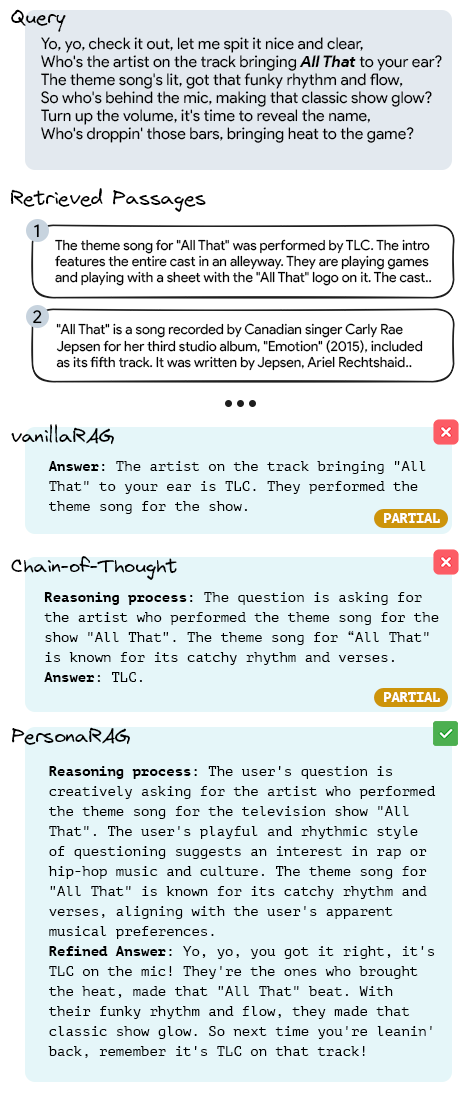}
   \caption{Illustrations of Various RAG Models. Vanilla RAG and Chain-of-Thought~\cite{YuZPMWY23} use passive learning, while PersonaRAG involves user-centric knowledge acquisition.}
   \label{fig:RAGmodels}
\end{figure}

\section{Introduction}
Large Language Models (LLMs) such as GPT-4~\cite{GPT4} and LLaMA 3~\cite{TouvronLIMLLR23} have significantly advanced the field of natural language processing (NLP) by demonstrating impressive performance across various tasks and exhibiting emergent abilities that push the boundaries of artificial intelligence~\cite{BrownMRSKDNSSAA20}. However, these models face challenges such as generating unreliable outputs due to issues like hallucination and outdated parametric memories~\cite{BangCLDSWLJYCDXF23}.

Retrieval-Augmented Generation (RAG) models have shown promise in addressing these issues by integrating externally retrieved information to support more effective performance on complex, knowledge-intensive tasks~\cite{LewisPPPKGKLYR020}. Despite these advancements, the deployment of RAG systems within broader AI frameworks continues to face significant challenges, particularly in handling noise and irrelevance in retrieved data~\cite{ChenLHS24}.

A key limitation of existing RAG systems is their inability to adapt outputs to users' specific informational and contextual needs. Personalized techniques in information retrieval, such as adaptive retrieval based on user interaction data and context-aware strategies, are increasingly recognized as essential for enhancing user interaction and satisfaction~\cite{TeevanDH17, SugiyamaHY04}. These methods aim to refine the retrieval process dynamically, tailoring it more closely to individual user profiles and situational contexts~\cite{AdomaviciusMRT11}.

The integration of agent-based systems with personalized RAG architectures presents a compelling avenue for research. Such systems utilize a multi-agent framework to simulate complex, adaptive interactions tailored to user-specific requirements~\cite{Wooldridge09}. By embedding intelligent, user-oriented agents within the RAG framework, these systems can evolve into more sophisticated tools that not only retrieve relevant information but also align it closely with the user's specific preferences and contexts in real-time. Importantly, the personalization strategy employed in these systems is fully transparent to the user, ensuring that the user is aware of how their information is being used to tailor the results.

In this study, we present PersonaRAG, an innovative methodology that extends traditional RAG frameworks by incorporating user-centric agents into the retrieval process. This approach addresses the previously mentioned limitations by promoting active engagement with retrieved content and utilizing dynamic, real-time user data to continuously refine and personalize interactions. PersonaRAG aims to enhance the precision and relevance of LLM outputs, adapting dynamically to user-specific needs while maintaining full transparency regarding the personalization process.

Our experiments, conducted using GPT-3.5, develop the PersonaRAG model and evaluate its performance across various question answering datasets. The results indicate that PersonaRAG achieves an improvement of over 5\% in accuracy compared to baseline models. Furthermore, PersonaRAG demonstrates an ability to adapt responses based on user profiles and information needs, enhancing the personalization of results. Additional analysis shows that the principles underlying PersonaRAG can be generalized to different LLM architectures, such as Llama 3 70b and Mixture of Experts (MoE) 8x7b~\cite{JiangSRMSB24}. These architectures benefit from the integration of external knowledge facilitated by PersonaRAG, with improvements exceeding 10\% in some cases. This evidence indicates that PersonaRAG not only contributes to the progress of RAG systems but also provides notable advantages for various LLM applications, signifying a meaningful step forward in the development of more intelligent and user-adapted information retrieval systems.

\section{Related Work}
Retrieval-Augmented Generation (RAG) systems have emerged as a significant advancement in natural language processing and machine learning, enhancing language models by integrating external knowledge bases to improve performance across various tasks, such as question answering, dialog understanding, and code generation~\cite{LewisPPPKGKLYR020, XuSC23}. These systems employ dense retrievers to pull relevant information, which the language model then uses to generate responses. However, the development of RAG systems and their integration within broader artificial intelligence frameworks is an ongoing area of research, with several challenges and opportunities for improvement.

Recent developments in RAG systems have focused on refining these models to better handle the noise and irrelevant information often retrieved during the process. ~\citet{XuSC23} addressed this issue by employing natural language inference models to select pertinent sentences, thereby enhancing the RAG's robustness. Additionally, advancements have been made in adaptively retrieving information, with systems like those proposed by ~\citet{JiangXGSLDYCN23} dynamically fetching passages that are most likely to improve generation accuracy.

Despite these improvements, RAG systems still face limitations, particularly in adapting their output to the user's specific profile, such as their information needs or intellectual knowledge. This limitation stems from the current design of most RAG systems, which do not typically incorporate user context or personalized information retrieval strategies~\cite{ZamaniC16a}. Consequently, there exists a gap between the general effectiveness of RAG systems and their applicability in personalized user experiences, where context and individual user preferences play a crucial role.

Personalization in information retrieval is increasingly recognized as essential for enhancing user interaction and satisfaction~\cite{GhorabZOW13}. Techniques such as user profiling, context-aware retrieval, and adaptive feedback mechanisms are commonly employed to tailor search results to individual users' needs. For instance, ~\citet{JeongBCH24} proposed adaptive retrieval strategies that dynamically adjust the retrieval process based on the complexity of the query and the user's historical interaction data. These personalized approaches not only improve user satisfaction but also increase the efficiency of information retrieval by reducing the time users spend sifting through irrelevant information.

The integration of personalized techniques with agent-based systems provides a promising pathway to augment the capabilities of RAG systems. Agent-based systems, particularly in the form of LLM-Based Multi-Agent Frameworks~\cite{LiZS23}, enable the simulation of complex interactions that can lead to more nuanced and contextually appropriate outputs. By incorporating multi-agent systems into RAG frameworks, there is potential for developing more robust and adaptive retrieval mechanisms that can handle a broader range of queries and generate more accurate responses, closely tailored to the specific needs and contexts of individual users.

In conclusion, while significant progress has been made in enhancing the effectiveness and personalization of RAG systems, ongoing research is crucial to address their existing limitations and expand their applications. The integration of personalized information retrieval and agent-based enhancements represents a promising avenue for further enhancing the adaptability and accuracy of RAG systems, potentially leading to intelligent information retrieval tailored to the specific needs of users.

\begin{figure*}[h]
   \centering
   \includegraphics[width=1\linewidth]{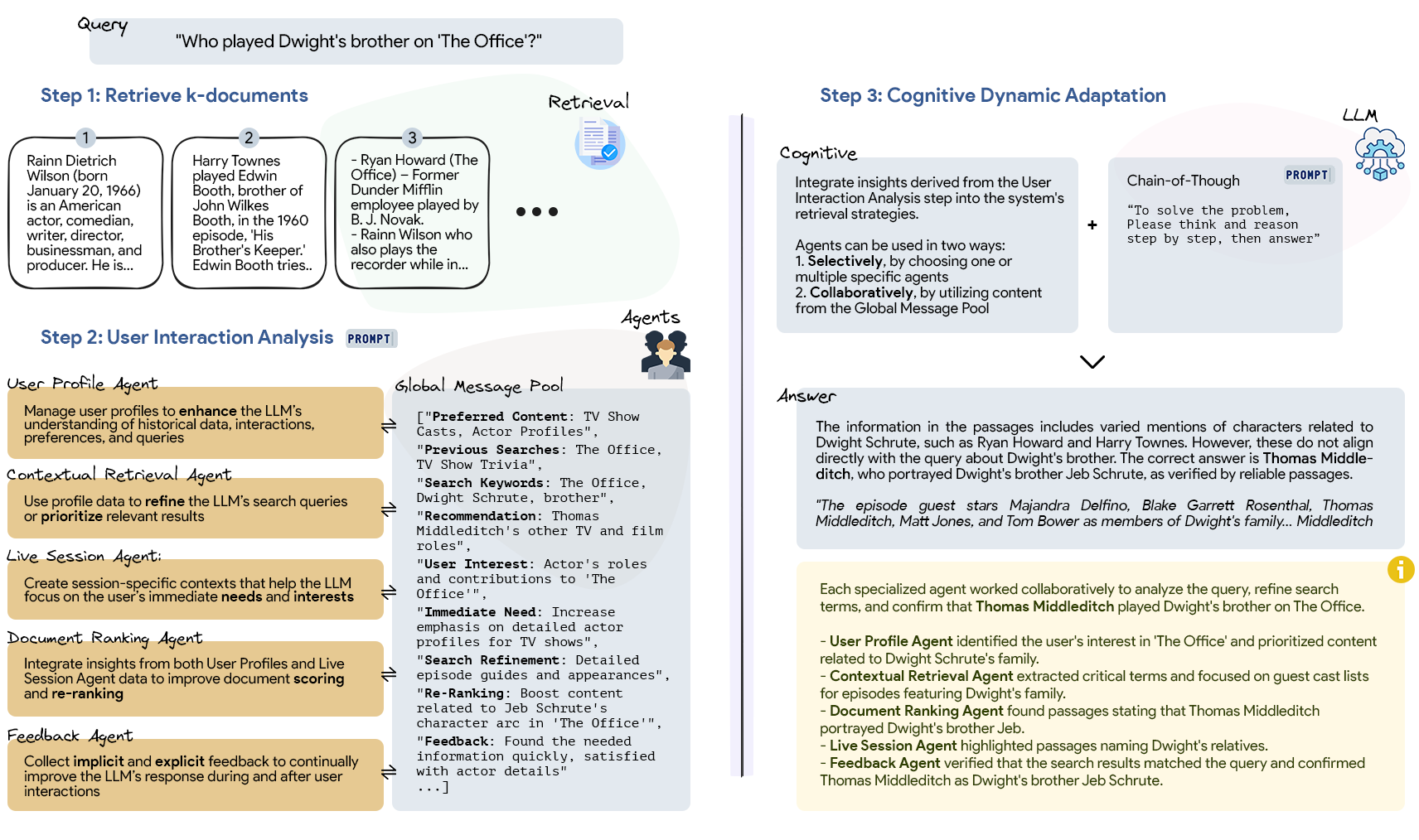}
   \caption{Overview of Our PersonaRAG Model showcasing the dynamic interaction among specialized agents within the system, facilitated by a global message pool for structured communication. The diagram illustrates the flow from user query input through various agents, including User Profile, Context Retrieval, Session Analysis, Document Ranking, and Feedback Agents, highlighting their contributions to real-time adaptation and personalized content generation by integrating live user data and feedback for continuous improvement and contextually relevant search experiences.}
   \label{fig:PersonaRAG}
\end{figure*}

\section{Methodology}
In this section, we present the methodology underlying our PersonaRAG approach, which aims to enhance the ability of Language Large Models (LLMs) to actively engage with, understand, and leverage user profile information for personalized content generation. We begin by discussing the fundamental concepts of Retrieval-Augmented Generation (RAG) models (Section ~\ref{sec:vanillaRAG}) and then introduce our PersonaRAG technique, which encourages LLMs to actively assimilate knowledge from live search sessions (Section ~\ref{sec:PersonaRAG}).

\subsection{Fundamentals of Retrieval-Augmented Generation (RAG) Models}
\label{sec:vanillaRAG}
State-of-the-art RAG models, as described in previous studies~\cite{GaoXGJPBDSGWW23, HuangH24, SiriwardhanaWKWRN23}, employ retrieval systems to identify a set of passages \( D = \{d_1, \ldots, d_n\} \) when given a query q. These passages are intended to enhance the generative capabilities of LLMs by providing them with contextually relevant information.

Early versions of RAG models typically employ a traditional retrieval-generation framework, in which the retrieved data set \( D = \{d_1, \ldots, d_n\} \) is directly fed into LLMs to generate responses to the query $ q $. However, these passages often contain irrelevant information, and the direct utilization approach in RAG has been shown to restrict the potential benefits of the RAG framework~\cite{0011LH024}. This limitation has sparked further discussion on how to improve LLMs by integrating retrieval results and outputs generated by the models themselves~\cite{WuWZ24}.

\subsection{PersonaRAG: RAG with User-Centric Agents}
\label{sec:PersonaRAG}
Drawing from the principles of adaptive learning and user-centered design, we develop a new PersonaRAG architecture to enable IR systems to dynamically learn from and adapt to user behavior in real-time. As shown in Figure ~\ref{fig:PersonaRAG}, PersonaRAG introduces a three-step pipeline: retrieval, user interaction analysis, and cognitive dynamic adaptation. Unlike traditional IR models that statically respond to queries, PersonaRAG focuses on leveraging live user data to continually refine its understanding and responses without the need for manual retraining.

\subsubsection{User Interaction Analysis}
To understand user behavior from live interactions, PersonaRAG treats the IR system as a cognitive structure capable of receiving, interpreting, and acting upon user feedback~\cite{AtkinsonS68}. Mimicking human learning behaviors, we establish four distinct agents within the system dedicated to analyzing user interactions from different perspectives: engagement tracking, preference analysis, context understanding, and feedback integration. These agents' roles are detailed in Section ~\ref{sec:AgentsRAG}.

\subsubsection{Cognitive Dynamic Adaptation}
Following adaptive learning principles, we employ a dynamic adaptation mechanism to assist the IR system in utilizing real-time user data for continuous improvement. This mechanism facilitates the integration of insights gained from User Interaction Analysis into the system's retrieval processes. Specifically, we prompt the system to adjust its query responses based on an initial understanding of the user's needs and refine these responses as more user data becomes available. This approach not only personalizes the search results but also helps in correcting any misalignments or errors in real-time.

\label{sec:AgentsRAG}
PersonaRAG employs a highly specialized agent architecture, with each agent focusing on a specific aspect of the information retrieval process. All agents utilize in-context learning, i.e., prompting, to perform their designated tasks. This role specialization allows for the efficient decomposition of complex user queries into manageable tasks~\cite{SharmaK20}. To foster this, we engage the IR system as five specialized agents to analyze user interactions based on retrieved data. At present, the focus is on the functionality and interaction of these agents rather than their individual performance metrics.

\paragraph{\textnormal{\textbf{User Profile Agent}}}
This component manages and updates user profile data, incorporating historical user interactions and preferences~\cite{Kacem17, SinghS19}. It monitors how users interact with search results, such as click-through rates and navigation paths. The User Profile Agent helps the system understand what captures user interest and leads to deeper engagement, enabling personalized search experiences.

\paragraph{\textnormal{\textbf{Contextual Retrieval Agent}}}
This agent is responsible for the initial retrieval of documents based on the user's current query. It accesses both a traditional search index and a more dynamic context-aware system that can consider broader aspects of the query environment. It utilizes user profile data to modify and refine search queries or to prioritize search results. For instance, if a user consistently engages more with certain types of documents or topics, the retrieval agent can boost those document types in the search results, ensuring that the most relevant information is presented to the user.

\paragraph{\textnormal{\textbf{Live Session Agent}}}
This agent analyzes the current session in real-time, observing user actions such as clicks, time spent on documents, modifications to the query, and any feedback provided. It creates a session-specific context model that captures the user's immediate needs and interests. The real-time data collected by this agent is used to adjust the ongoing session, potentially re-ranking search results or suggesting new queries based on the user's behavior and preferences. Additionally, the Live Session Agent updates the user profile with new insights gleaned from the session, allowing for a more personalized and efficient search experience in future interactions.

\paragraph{\textnormal{\textbf{Document Ranking Agent}}}
This agent is responsible for re-ranking the documents retrieved by the Contextual Retrieval Agent. It integrates insights from both the User Profile Agent and the Live Session Agent to score and order the documents more effectively. By considering the user's historical preferences and their current session behavior, the Document Ranking Agent ensures that the most relevant and valuable documents are presented to the user in a prioritized manner. This agent continuously adapts its ranking algorithms based on the feedback received from the user and the insights provided by the other agents in the system.

\paragraph{\textnormal{\textbf{Feedback Agent}}}
This agent gathers implicit and explicit feedback during and after user interactions. Implicit feedback includes behavioral data like time spent on documents, click counts, and navigation patterns. Explicit feedback involves direct user input on document relevance and quality, collected through ratings, surveys, or comments. The agent uses this information to train and refine models for other agents, particularly the Document Ranking Agent. This process enhances the system's ability to anticipate user needs and deliver relevant documents based on accumulated feedback and insights.


\smallskip
By dynamically integrating insights from the User Profile Agent, Contextual Retrieval Agent, Live Session Agent, Document Ranking Agent, and Feedback Agent into the IR processes, PersonaRAG not only adapts to immediate user needs but also evolves over time to better anticipate and meet user expectations. This multi-agent approach enables PersonaRAG to embody a truly adaptive and user-focused information retrieval system, leveraging specialized agents to analyze user interactions from different behavioral perspectives and deliver highly personalized and contextually relevant search experiences. The inclusion of the Document Ranking Agent ensures that the most pertinent documents are identified and presented to users, further enhancing the system's ability to effectively satisfy user information needs.

\subsection{PersonaRAG Operational Workflow}
The PersonaRAG framework employs a structured workflow that allows for sequential and parallel processing of tasks, ensuring clarity and consistency in communication between agents through well-defined data structures and protocols~\cite{HongZCCWZWYLZRX23}. The process involves the User Profile Agent, Contextual Retrieval Agent, Live Session Agent, Document Ranking Agent, and Feedback Agent working together to refine search queries, prioritize relevant results, and improve document scoring and re-ranking based on user profile, session-specific contexts, and feedback.

PersonaRAG's modular design allows for flexibility in the system setup, enabling researchers to focus on the most relevant aspects of the user's profile, session, and feedback data. Agents work collaboratively by utilizing content from the Global Message Pool, which serves as a central hub for inter-agent communication~\cite{HongZCCWZWYLZRX23}, eliminating inefficiencies and enabling agents to access or update information as required.

The Feedback Agent collects and analyzes implicit and explicit user feedback to generate insights into the effectiveness of retrieval strategies and document relevance. This feedback is used to make dynamic adjustments to the system, refining retrieval methods and altering the weighting of user profile factors. Through this iterative process, PersonaRAG continuously adapts and improves its performance, enhancing the accuracy and user satisfaction of the retrieval results~\cite{LimbuCPM06}.

\section{Experimental Setups}
In this section, we present the experimental setup employed in our study, including the datasets, baseline models, evaluation metrics, and implementation details. We also provide an overview of the prompts used in our experiments.

\subsection{Datasets}
Our experiments are conducted on three widely used single-hop benchmark datasets in the field of Information Retrieval (IR): NaturalQuestions (NQ)~\cite{KwiatkowskiPRCP19}, TriviaQA~\cite{JoshiCWZ17}, and WebQuestions (WebQ)~\cite{BerantCFL13}. NQ is a well-known dataset in Natural Language Understanding (NLU), consisting of structured questions and corresponding Wikipedia pages annotated with long and short answers. TriviaQA comprises question-answer pairs collected from trivia and quiz-league websites, while WebQ consists of questions selected using the Google Suggest API, with answers being entities in Freebase.

Table ~\ref{tab:datasets} summarizes the datasets used in our initial study. Due to the high cost of using language models and the large number of API calls required, we randomly sampled 500 questions from each raw dataset to create more manageable subsets for our experiments. While this sampling approach limits the scope of our study, it allows us to conduct an initial investigation into the performance of different RAG systems on these datasets. We acknowledge that future work with larger sample sizes and more comprehensive experiments will be necessary to draw definitive conclusions. Nonetheless, we believe this preliminary study provides valuable insights into the relative strengths and weaknesses of the tested RAG approaches.

\begin{table}[ht]
\centering 
\begin{tabular}{lccc}
\toprule
\textbf{Dataset} & \textbf{\#Query} & \textbf{\#Corpus} & \textbf{Sampling Rate} \\ 
\midrule
NQ        & 8,757   & 79,168  & 5.7\% \\ 
TriviaQA  & 8,837   & 78,785  & 5.7\% \\ 
WebQ      & 2,032   & 3,417   & 24.6\% \\ 
\bottomrule 
\end{tabular}
\caption{Summary of datasets. Each dataset consists of randomly sampled 500 questions from the raw dataset.} 
\label{tab:datasets} 
\end{table}

\subsection{Models}
We compare PersonaRAG with several baseline models, including prompt learning and RAG models. The prompt templates used in user interaction analysis and dynamic adaptation are presented in Section ~\ref{sec:implementation}. Initially, the question-answering (QA) instruction is fed to ChatGPT to conduct the vanilla answer generation model. Following the work of ~\citet{Wei0SBIXCLZ22}, the Chain-of-Thought model is implemented, which generates question rationale results to produce the final results. Additionally, the Guideline model serves as a baseline, generating problem-solving steps and guiding Language Models (LLMs) to generate the answer.

For the RAG-based baselines, two models are implemented: vanilla RAG and Chain-of-Thought, which include utilizing raw retrieved passages (CoT with Passage) and refining the passages as notes (CoT with Note). The vanilla RAG model directly feeds the top-ranked passages to the LLM. The Chain-of-Note model~\cite{YuZPMWY23} is also implemented, which refines and summarizes the retrieved passages for generation. Inspired by Self-RAG~\citet{AsaiWWSH23}, the Self-Rerank model is conducted, which filters out unrelated contents without fine-tuning LLMs.

\subsection{Evaluation Metrics}
When evaluating adaptive models, it is crucial to consider both task performance and user-centric adaptability simultaneously, along with their trade-offs. Therefore, the results are reported using different metrics, some of which measure effectiveness and others measure efficiency.

For effectiveness, accuracy is used, following the standard evaluation protocol in the field of Information Retrieval (IR)~\cite{MallenAZDKH23, BaekJKPH23, AsaiWWSH23}. Accuracy assesses whether the predicted answer contains the ground-truth answer. Both the outputs of the Language Learning Model (LLM) and golden answers are converted to lowercase, and string matching (StringEM) is performed between each golden answer and the model prediction to calculate accuracy.

To evaluate user-centric adaptability, the BLEU-2 score is measured to assess the text similarity between different RAG and baseline setups and how well the generated answers resemble each other. This metric provides insights into the system's ability to generate consistent and coherent responses across various configurations. Additionally, the average sentence length and the average number of syllables of the answers from different RAG setups are reported as a post-hoc analysis. These measures validate whether the RAG system effectively adjusts its responses based on user knowledge levels, ensuring that the generated answers are tailored to the user's understanding and expertise.

Combining these evaluation strategies provides a comprehensive view of both the effectiveness and user-centric adaptability of the RAG system. The accuracy metric ensures that the system generates correct answers, while the BLEU-2 score and post-hoc analysis of sentence length and syllable count confirm the system's ability to adapt to user knowledge levels. As the understanding of user needs and system capabilities evolves, it is essential to continuously refine these metrics to maintain the RAG system's effectiveness in delivering personalized, context-aware responses that cater to the diverse requirements of users in the field of IR.

\subsection{Implementation Details}
\label{sec:implementation}
For a fair comparison and following the work of ~\citet{MallenAZDKH23} and ~\citet{TrivediBKS23}, the same retriever, a term-based sparse retrieval model known as BM25~\cite{RobertsonWJHG94}, is used across all different models. The retrieval model is implemented using the OpenMatch toolkit~\cite{YuLX023}. For the external document corpus, the KILT-Wikipedia corpus preprocessed by ~\citet{PetroniPFLYCTJK21} is used, and the top-k relevant documents are retrieved.

Regarding the LLMs used to generate answers, the Llama 3 model instruct (ref) with 70b parameters, Mixture of Experts (MoE) 8x7b (ref), and the GPT-3.5 model \texttt{(gpt-3.5-turbo-0125)} are employed. For the retrieval-augmented LLM design, the implementation details from ~\citet{TrivediBKS23} are followed, which include input prompts, instructions, and the number of test samples for evaluation (e.g., 500 samples per dataset).

\subsection{Prompts Used in PersonaRAG}
This subsection presents the prompt templates employed in the construction of the PersonaRAG model. The prompts utilized in the User Interaction Analysis and Cognitive Dynamic Adaptation components are detailed below. The prompt templates used by the baseline models are available in the project repository~\footnote{\url{https://anonymous.4open.science/r/PersonaRAG-F423}}. In the templates, \texttt{\{question\}} represents the input question, \texttt{\{global\_memory\}} the Global Message Pool, while \texttt{\{passages\}} denotes the retrieved passages. Additionally, \texttt{\{cot\_answer\}} is populated with the output generated by the Chain-of-Thought model.

The placeholder \texttt{\{user\_profile\_answer\}} is filled with the response produced by the User Profile agent model. Respectively, \texttt{\{contextual\_answer\}} corresponds to the Contextual Retrieval agent model, \texttt{\{live\_session\_answer\}} to the Live Session agent model, \texttt{\{document\_ranking\_answer\}} to the Document Ranking agent model, and \texttt{\{feedback\_answer\}} to the Feedback agent model.

\subsubsection{Prompts Used in User Interaction Analysis}
\paragraph{\textnormal{\textbf{User Profile Agent}}}
\begin{promptbox}
Your task is to help the User Profile Agent improve its understanding of user preferences based on ranked document lists and the shared global memory pool. \\ 

\textbf{Question:} \{question\} \\
\textbf{Passages:} \{passages\} \\ 
\textbf{Global Memory:} \{global\_memory\} \\ 

\textbf{Task Description:} \\
From the provided passages and global memory pool, analyze clues about the user's search preferences. Look for themes, types of documents, and navigation behaviors that reveal user interest. Use these insights to recommend how the User Profile Agent can refine and expand the user profile to deliver better-personalized results.
\end{promptbox}

\paragraph{\textnormal{\textbf{Contextual Retrieval Agent}}}
\begin{promptbox}
You are a search technology expert guiding the Contextual Retrieval Agent to deliver context-aware document retrieval. \\ 

\textbf{Question:} \{question\} \\
\textbf{Passages:} \{passages\} \\ 
\textbf{Global Memory:} \{global\_memory\} \\ 

\textbf{Task Description:} \\
Using the global memory pool and the retrieved passages, identify strategies to refine document retrieval. Highlight how user preferences, immediate needs, and global insights can be leveraged to adjust search queries and prioritize results that align with the user's interests. Ensure the Contextual Retrieval Agent uses this shared information to deliver more relevant and valuable results.
\end{promptbox}

\paragraph{\textnormal{\textbf{Live Session Agent}}}
\begin{promptbox}
Your expertise in session analysis is required to assist the Live Session Agent in dynamically adjusting results. \\ 

\textbf{Question:} \{question\} \\
\textbf{Passages:} \{passages\} \\ 
\textbf{Global Memory:} \{global\_memory\} \\ 

\textbf{Task Description:} \\
Examine the retrieved passages and information in the global memory pool. Determine how the Live Session Agent can use this data to refine its understanding of the user's immediate needs. Suggest ways to dynamically adjust search results or recommend new queries in real-time, ensuring that session adjustments align with user preferences and goals.
\end{promptbox}

\paragraph{\textnormal{\textbf{Document Ranking Agent}}}
\begin{promptbox}
Your task is to help the Document Ranking Agent prioritize documents for better ranking. \\ 

\textbf{Question:} \{question\} \\
\textbf{Passages:} \{passages\} \\ 
\textbf{Global Memory:} \{global\_memory\} \\ 

\textbf{Task Description:} \\
Analyze the retrieved passages and global memory pool to identify ways to rank documents effectively. Focus on combining historical user preferences, immediate needs, and session behavior to refine ranking algorithms. Your insights should ensure that documents presented by the Document Ranking Agent are prioritized to match user interests and search context.
\end{promptbox}

\paragraph{\textnormal{\textbf{Feedback Agent}}}
\begin{promptbox}
You are an expert in feedback collection and analysis, guiding the Feedback Agent to gather and utilize user insights. \\ 

\textbf{Question:} \{question\} \\
\textbf{Passages:} \{passages\} \\ 
\textbf{Global Memory:} \{global\_memory\} \\ 

\textbf{Task Description:} \\
Using the retrieved passages and global memory pool, identify methods for collecting implicit and explicit user feedback. Suggest ways to refine feedback mechanisms to align with user preferences, such as ratings, surveys, or behavioral data. Your recommendations should guide the Feedback Agent in updating other agents' models for more personalized and relevant results.
\end{promptbox}

\paragraph{\textnormal{\textbf{Global Message Pool}}}
\begin{promptbox}
You are responsible for maintaining and enriching the Global Message Pool, serving as a central hub for inter-agent communication. \\ 

\textbf{Question:} \{question\} \\
\textbf{Agent Responses:} \{agent\_responses\} \\ 
\textbf{Existing Global Memory:} \{global\_memory\} \\ 

\textbf{Task Description:} \\
Using the responses from individual agents and the existing global memory, consolidate key insights into a shared repository. Your goal is to organize a comprehensive message pool that includes agent-specific findings, historical user preferences, session-specific behaviors, search queries, and user feedback. This structure should provide all agents with meaningful data points and strategic recommendations, reducing redundant communication and improving the system's overall efficiency.
\end{promptbox}

\subsubsection{Prompts Used in Cognitive Dynamic Adaptation}
\paragraph{\textnormal{\textbf{Chain-of-Thought}}}
\begin{promptbox}
To solve the problem, Please think and reason step by step, then answer. \\ 

\textbf{Question:} \{question\} \\
\textbf{Passages:} \{passages\} \\ 
\textbf{Reasoning process:} \\
1. Read the given question and passages to gather relevant information. \\
2. Write reading notes summarizing the key points from these passages. \\
3. Discuss the relevance of the given question and passages. \\
4. If some passages are relevant to the given question, provide a brief answer based on the passages. \\
5. If no passage is relevant, directly provide the answer without considering the passages. \\

\textbf{Answer:}
\end{promptbox}

\paragraph{\textnormal{\textbf{Cognitive Agent}}}
\begin{promptbox}
Your task is to help the Cognitive Agent enhance its understanding of user insights to continuously improve the system's responses. \\ 

\textbf{Question:} \{question\} \\
\textbf{Initial Response:} \{cot\_answer\} \\ 
\textbf{User Insights from Interaction Analysis:} \\ 
User Profile Agent: \{user\_profile\_answer\}, \\ 
Contextual Retrieval Agent: \{contextual\_answer\}, \\ 
Live Session Agent: \{live\_session\_answer\}, \\ 
Document Ranking Agent: \{document\_ranking\_answer\}, \\ 
Feedback Agent: \{feedback\_answer\} \\ 

\textbf{Task Description:} \\
Verify the reasoning process in the initial response for errors or misalignments. Use insights from user interaction analysis to refine this response, correcting any inaccuracies and enhancing the query answers based on user profile. Ensure that your refined response aligns more closely with the user's immediate needs and incorporates foundational or advanced knowledge from other sources. \\

\textbf{Answer:}
\end{promptbox}

\section{Experimental Results and Analyses}
In this section, we show the overall experimental results and offer in-depth analyses of our method.

\begin{table*}[h]
    \centering
    \begin{tabular}{|l|l|c|c|c|c|c|c|}
        \hline
        \textbf{Method} & \textbf{Setting} & \multicolumn{3}{c|}{\textbf{Top-3}} & \multicolumn{3}{c|}{\textbf{Top-5}} \\ \hline
        & & \textbf{WebQ} & \textbf{TriviaQA} & \textbf{NQ} & \textbf{WebQ} & \textbf{TriviaQA} & \textbf{NQ} \\ \hline
        \multirow{3}{*}{w/o RAG} & \texttt{gpt-3.5-turbo-0125} & 59.61 & 97.36 & 43.90 & 62.43 & 97.36 & 41.46 \\ 
                                 & Guideline & 36.53 & 42.10 & 17.07 & 47.21 & 36.84 & 21.95 \\ \hline
        \multicolumn{2}{|l|}{vanillaRAG} & 38.46 & 78.94 & 36.58 & 50.14 & 81.57 & 41.46 \\ \hline
        \multirow{3}{*}{Self-Refined} & Chain-of-Thought (CoT)  & 57.69 & 89.47 & 39.02 & 67.51 & 89.47 & 41.46 \\ 
                                          & Chain-of-Note (CoN) & 57.17 & 81.57 & 48.78 & 65.15 & 92.10 & 48.78 \\ 
                                          & Self-Rerank (SR) & 32.63 & 81.57 & 43.90 & 40.26 & 84.21 & 51.21 \\ \hline
        \multicolumn{2}{|l|}{PersonaRAG} & \textbf{63.46} & \textbf{94.73} & \textbf{49.02} & \textbf{67.50} & \textbf{89.47} & \textbf{48.78} \\ \hline
    \end{tabular}
    \caption{Overall Accuracy Performance Comparison Using Top-3 and Top-5 Passages. PersonaRAG results are reported in \textbf{bold}.}
    \label{tab:performance_comparison}
\end{table*}


\subsection{Main Results}
Table ~\ref{tab:performance_comparison} summarizes the primary findings for PersonaRAG across various single-hop question answering datasets. The approach was evaluated against multiple baseline models, including large language models (LLMs) without retrieval-augmented generation (RAG), the conventional RAG model, and self-refined variants, such as utilizing raw retrieved passages (CoT with Passage) or refining passages into notes (CoT with Note).

PersonaRAG demonstrated superior performance compared to most of the baseline models, achieving significant improvements over the conventional RAG (i.e., vanillaRAG) of over 10\%, particularly on the WebQ dataset. It also consistently outperformed the ChatGPT-3.5 model, except on TriviaQA, which we suspect is part of the model's training dataset. These results suggest PersonaRAG's capability to guide LLMs in extracting relevant information through active learning techniques.

Specifically, the performance of RAG models was assessed using the top 3 and 5 ranked passages. While other RAG models generally benefited from more passages, PersonaRAG maintained consistent performance with either 3 or 5 passages, suggesting that 3 passages were adequate for generating accurate answers. PersonaRAG agents played a crucial role in efficiently extracting the necessary information regarding the user's information need to achieve these improvements.

Furthermore, on the WebQ dataset, PersonaRAG achieved accuracy scores of 63.46\% and 67.50\% using Top-3 and Top-5 passages, respectively, surpassing the vanillaRAG model by 25\% and 17.36\%, and nearly all other baseline models (except for Chain-of-Thought using Top-5, which performed equally). On the NQ dataset, PersonaRAG maintained similarly robust performance with scores of 49.02\% and 48.78\%, outperforming all baselines (except for Chain-of-Thought and Self-Rerank (SR) using Top-5). This pattern was further validated by experiments on other datasets, with results showing that PersonaRAG consistently outperforms conventional RAG models with the capability of providing an answer tailored to the user's interaction and information need. The comprehensive understanding it provides contributes to the generation of accurate and user-centric answers across various question complexities.

\subsection{Comparative Analysis of RAG Configurations}
\begin{figure}[h]
   \centering
   \begin{subfigure}[b]{1\linewidth}
       \centering
       \includegraphics[width=\linewidth]{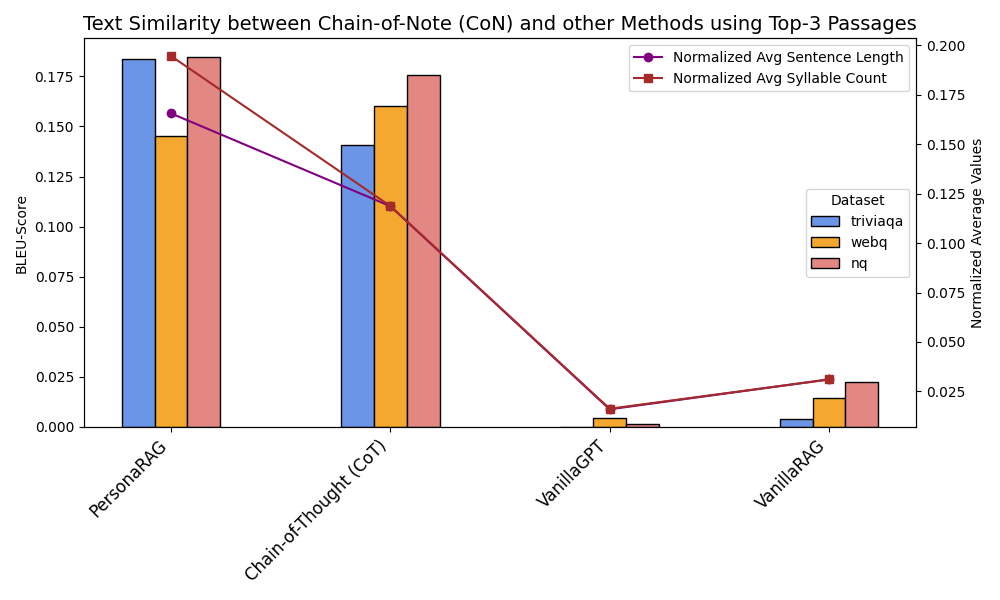}
       \caption{Text Similarity for Top-3 Passages}
   \end{subfigure}

   \begin{subfigure}[b]{1\linewidth}
       \centering
       \includegraphics[width=\linewidth]{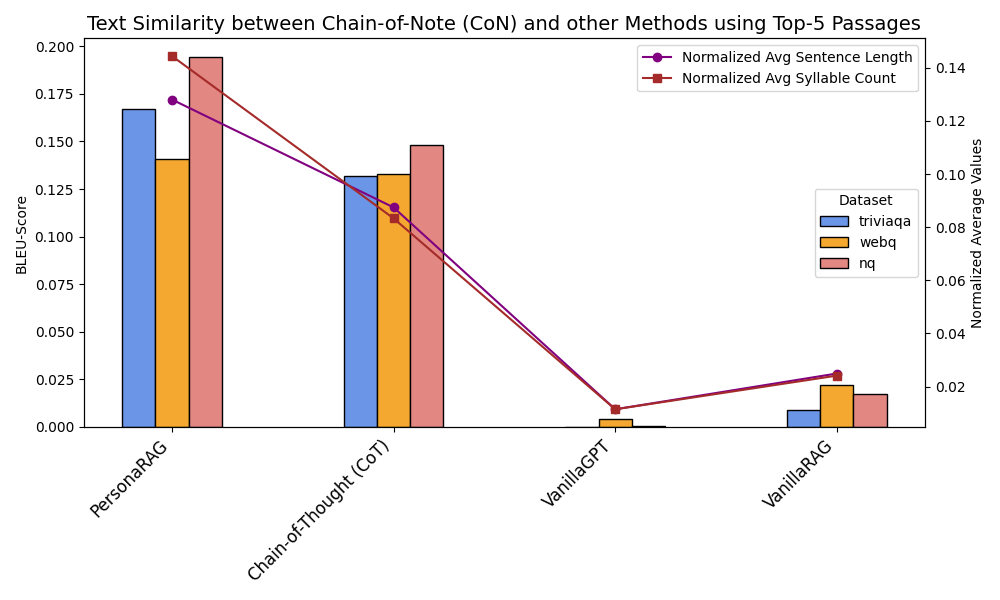}
       \caption{Text Similarity for Top-5 Passages}
   \end{subfigure}

   \caption{Text Similarity between Chain-of-Note (CoN) and Other Methods Using BLEU-2 Score for Evaluation, with Normalized Average Sentence Length and Average Syllable Count.}
   \label{fig:RAGconfigs}
\end{figure}

Further experiments explored PersonaRAG's adaptive capabilities (Figure ~\ref{fig:RAGconfigs}). BLEU-2 scores compared outputs from Chain-of-Note (consistently best outside PersonaRAG) with other methods. PersonaRAG showed higher similarity scores, indicating its ability to generate responses that address user needs rather than just summarizing input. Additionally, PersonaRAG provides personalized answers tailored to user profiles, extending beyond mere information provision.

The Chain-of-Note approach demonstrated comparable performance to the Chain-of-Thought approach, implying that both techniques effectively extract pertinent information from the retrieved passages and adapt it to align with the user's information need.



In contrast, vanillaGPT and vanillaRAG outputs differed significantly from the Chain-of-Note approach, indicating that counterfactual cognition often leads to diverse outcomes rather than focusing solely on query-relevant content. This suggests LLMs can construct knowledge from multiple perspectives and customize responses based on user understanding.

Post-hoc analyses of average sentence length and syllable count across RAG configurations provided insights into the system's ability to adapt responses to user comprehension levels. These observations highlight PersonaRAG's capacity to synthesize knowledge from various perspectives and tailor responses to different levels of user expertise.

\begin{table}[h]
    \centering
    \begin{tabular}{|p{2.5cm}|>{\centering\arraybackslash}p{1.25cm}|>{\centering\arraybackslash}p{1.25cm}|>{\centering\arraybackslash}p{1.25cm}|}
        \hline
        \textbf{Method} & \textbf{WebQ} & \textbf{TriviaQA} & \textbf{NQ} \\ \hline
        \multicolumn{4}{|l|}{\textbf{LLaMA3-70B}} \\ \hline
        w/o RAG & 45.25 & 82.17 & 38.95 \\ \hline
        vanillaRAG & 55.14 & 85.02 & 40.37 \\ \hline
        Chain-of-Thought & 60.52 & 88.72 & 45.10 \\ 
        Chain-of-Note & 62.67 & 89.37 & 48.25 \\ 
        Self-Rerank & 54.25 & 84.50 & 47.77 \\ \hline
        PersonaRAG & \textbf{66.09} & \textbf{92.12} & \textbf{50.85} \\ \hline
        \multicolumn{4}{|l|}{\textbf{MoE-8x7b}} \\ \hline
        w/o RAG & 38.24 & 75.82 & 34.26 \\ \hline
        vanillaRAG & 48.44 & 80.25 & 38.50 \\ \hline
        Chain-of-Thought & 54.12 & 85.46 & 42.37 \\ 
        Chain-of-Note & 55.98 & 87.55 & 45.14 \\ 
        Self-Rerank & 52.50 & 83.04 & 44.96 \\ \hline
        PersonaRAG & \textbf{61.35} & \textbf{90.30} & \textbf{49.27} \\ \hline
    \end{tabular}
    \caption{Overall Accuracy Performance Comparison for Top-5 Passages using LLaMA3-70B and MoE-8x7b.}
    \label{tab:accuracy_comparison}
\end{table}

\subsection{Analysis on Generalization Ability}
This experiment evaluates the quality of knowledge construction using different large language models (LLMs). As illustrated in Table ~\ref{tab:accuracy_comparison}, the PersonaRAG outcomes are used to prompt open-source LLMs, specifically LLaMA3-70B and MoE-8x7b, to generate accurate answers.

Compared to LLMs without retrieval-augmented generation (w/o RAG), vanilla RAG and Chain-of-Note often exhibit lower performance. This result suggests that retrieved passages can act as noise, adversely affecting model performance even after refinement through note generation. One primary reason for this behavior is that both LLaMA3-70B and MoE-8x7b struggle to efficiently analyze and identify relevant knowledge due to limitations in their processing capacities.

In contrast, the PersonaRAG method provides notable performance improvements: over 8\% for LLaMA3-70B and more than 10\% for MoE-8x7b across all datasets, underscoring its effectiveness. The PersonaRAG methodology distinguishes itself from the Chain-of-Note approach by offering a cognitive framework that connects retrieved passages with prior knowledge. This framework models the instructor's (GPT-3.5) reasoning process, guiding the student models (LLaMA3-70B and MoE-8x7b) to better understand knowledge retrieved from passages. The results demonstrate that the LLMs are capable of selecting appropriate passages to build more accurate responses, highlighting the benefits of the PersonaRAG approach for improving generalization.

\begin{table}[h]
\begin{tabular}{>{\raggedright\arraybackslash}p{8cm}}
\hline
\textbf{Question:} Who stole the Mona Lisa from the Louvre in 1911? \\ \textbf{Answer:} "Vincenzo Peruggia" \\ \hline
\textbf{Chain-of-Thought:} \\
1. Read the given question and passages to gather relevant information.  \\ 
2. Write reading notes summarizing the key points.  \\ 
3. Discuss the relevance of the given question and passages.  \\ 
4. If some passages are relevant to the given question, provide a brief answer based on the passages.  \\ 
5. If no passage is relevant, directly provide the answer without considering the passages. \\ \hline 
\textbf{Step 1: Passage Retrieval} \\ \hline
1. The Mona Lisa was stolen from the Louvre Museum while the museum was closed for cleaning. Witnesses reported that a tall, stout individual was carrying what appeared to be a large panel covered with a horse blanket. Two years later, Vincenzo Peruggia was arrested in Florence.  \\
2. On August 22, 1911, the Mona Lisa was stolen from the Louvre. Vincenzo Peruggia claimed he stole the painting to return it to Italy.  \\
3. In 1911, Vincenzo Peruggia, a Louvre employee, stole the Mona Lisa.  \\ \hline
\textbf{Step 2: User Interaction Analysis} \\ \hline
User Insights from Interaction Analysis:  \\
- User Profile Agent: The user is interested in art heists and historical mysteries.  \\
- Contextual Retrieval Agent: The user frequently searches for art history topics.  \\
- Live Session Agent: The user reads about unsolved mysteries and historic crimes.  \\
- Document Ranking Agent: The user favors detailed, chronological accounts.  \\
- Feedback Agent: The user interacts with articles about famous art heists.  \\
\hline
\textbf{Step 3: Cognitive Dynamic Adaptation} \\ \hline
The user has a consistent interest in famous art heists and historical mysteries. They are drawn to unsolved mysteries and historic crimes. They favor detailed, chronological accounts, so the final response should highlight theft motives, the method, and the investigation leading to the arrest, providing comprehensive and authoritative sources to satisfy their engagement with articles on art heists. \\
\textbf{Answer:} Vincenzo Peruggia, a Louvre employee, stole the Mona Lisa from the Louvre Museum on August 21, 1911. He claimed he stole the painting to return it to Italy, and the theft remained unsolved for two years before he was arrested in Florence in December 1913. The painting was later recovered and returned to the Louvre.  \\ \hline

\end{tabular}

    \caption{PersonaRAG Case Study.}
    \label{tab:usecase}
\end{table}

\subsection{Case Study}
Finally, we randomly sample one case in Table ~\label{ref:usecase} to demonstrate the effectiveness of PersonaRAG.

The user interaction analysis mechanism effectively generates comprehensive results by integrating foundational and advanced insights from user data. Retrieved passages provide critical clues for answering questions, while agent analyses summarize and illustrate the applicability of external information to user queries. The cognitive dynamic adaptation module refines initial chain-of-thought responses using these insights, generating accurate answers. For example, including knowledge about the "theft of the Mona Lisa in 1911," "Vincenzo Peruggia," and "Florence" enhances the reasoning process's precision and detail. This demonstrates PersonaRAG's effectiveness in helping IR agents combine external knowledge with intrinsic user data to produce well-informed responses.

\section{Conclusion}
This paper proposes PersonaRAG, which constructs the retrieval-augmentation architecture incorporating user interaction analysis and cognitive dynamic adaptation. PersonaRAG builds the user interaction agents and dynamic cognitive mechanisms to facilitate the understanding of user needs and interests and enhance the system capabilities to deliver personalized, context-aware responses with the intrinsic cognition of LLMs.

Furthermore, PersonaRAG demonstrates effectiveness in leveraging external knowledge and adapting responses based on user profiles, knowledge levels, and information needs to support LLMs in generation tasks without fine-tuning. However, this approach requires multiple calls to the LLM's API, which can introduce additional time latency and increase API calling costs when addressing questions. The process involves constructing the initial Chain-of-Thought, processing the User Interaction Agents results, and executing the Cognitive Dynamic Adaptation to generate the final answer. Furthermore, the inputs to LLMs in this approach tend to be lengthy due to the inclusion of extensive retrieved passages and the incorporation of user needs, interests, and profile construction results. These factors can impact the efficiency and cost-effectiveness of the PersonaRAG approach in practical applications of Information Retrieval (IR) systems. 

Future research will aim to optimize the process by reducing API calls and developing concise representations of user profiles and retrieved information without compromising response quality. We also plan to explore more user-centric agents to better capture writing styles and characteristics of RAG users/searchers. This will enhance the system's ability to understand and adapt to individual preferences, improving personalization and relevance in IR tasks.


\begin{acks}
This work has received funding from the European Union's Horizon Europe research and innovation program under grant agreement No 101070014 (OpenWebSearch.EU, \url{https://doi.org/10.3030/101070014}).
\end{acks}

\bibliographystyle{ACM-Reference-Format}
\bibliography{sample-base}

\end{document}